\documentclass[]{aa}
\usepackage{epsfig}
\usepackage{graphics}
\begin{document}
\thesaurus{06(02.01.1; 02.09.1; 06.16.1; 06.24.1; 06.03.2; 06.06.3)}
\title{Electron Firehose instability and acceleration of electrons in
solar flares} 
\author{Gunnar Paesold\inst{1,2} \and Arnold O. Benz\inst{1}}
\offprints{G. Paesold}
\mail{gpaesold@astro.phys.ethz.ch}
\institute{Institute of Astronomy, ETH-Zentrum, CH-8092 Zurich,
Switzerland \and Paul Scherrer Institute, W\"urenlingen und Villigen,
CH-5232 Villigen PSI, Switzerland}
\date{A \& A 1999, 351, 741-746 }
\titlerunning{Electron Firehose instability in solar flares}
\authorrunning{G. Paesold \& A.O. Benz}
\maketitle
\begin{abstract}
An electron distribution with a temperature
anisotropy $T_\parallel/T_\perp>1$ can lead to the Electron Firehose
instability (Here $\parallel$ and $\perp$ denote directions relative to
the background magnetic field ${\bf B}_0$). Since possible particle
acceleration mechanisms in solar flares exhibit a preference of
energizing particles in parallel direction, such an
anisotropy is expected during the impulsive phase of a flare. The
properties of the excited waves and the thresholds for instability are
investigated by using linearized kinetic theory. These thresholds were
connected to the pre-flare plasma parameters by assuming an acceleration model
acting exclusively in parallel direction. For usually assumed
pre-flare plasma conditions the electrons become unstable during the
acceleration process and lefthand circularly polarized waves with
frequencies of about $\sim|\Omega_p|$ are excited at parallel
propagation. Indications have been found, that the largest growth
rates occur at oblique propagation and the according frequencies lie
well above the proton gyrofrequency.  
\end{abstract} 
\section{Introduction}
Particle acceleration is a phenomenon occurring at many different sites
throughout the universe. An important example of particle acceleration
are solar flares,  
offering a wide range of observations that allow one to probe electron and ion 
acceleration. It is now widely accepted that the hard X-ray emission observed 
by various spacecraft reflects the energization of almost all electrons 
in the flaring plasma to energies up to $\sim25\mathrm{keV}$. These 
observations and the observed magnetic fields encompassing the solar
flare suggest that most of the dissipated energy is released by
restructuring the magnetic field, e.g. magnetic reconnection events.

During the impulsive phase of the flare, when the most powerful energization 
takes place, electrons must be accelerated to mean energies of about $\sim
25\mathrm{keV}$ at a rate of about $10^{36}$ electrons per second in
order to sustain the observed intensity of the hard X-ray
bursts. Taking the impulsive phase of a flare to last about
$10\mathrm{s}$ and assuming an electron density of about
$10^{10}\mathrm{cm}^{-3}$ (Moore \&Fung  
\cite{moorefung1972}; Vaiana \& Rosner \cite{vaianarosner1978}), the 
bulk energization must process a coronal volume of at least 
$10^{27}\mathrm{cm}^3$. Thus energization must affect a large fraction
of the electron population in the flaring region. 

In view of this background we want to briefly describe the processes which 
may be responsible for particle acceleration in solar flares. For a
detailed review of possible acceleration processes in impulsive solar
flares see e.g. Miller et al. (\cite{milleretal1997}) and Cargill
(\cite{cargill1999}). 
  
1) {\sl Shock Acceleration:} There are two types of shock acceleration. 
The one referred to as 'shock drift acceleration' involves the shock 
electric field that reflects and accelerates the particles moving
along the shock surface. Since this mechanism is only effective when
the shock normal approaches an angle of $90^\circ$ to the background
magnetic field, either the  
gained energy or the particle flux is very limited. It 
seems to be unlikely that this mechanism is responsible for the large 
number of accelerated particles in solar flares. The second kind of shock 
acceleration is called 'diffusive shock acceleration'.  In this process the
particles cross the shock-front several times, interacting with 
scattering centers on both sides of the shock. In the rest frame of the 
shock these centers approach each other and the particles systematically 
gain energy. This kind of acceleration process requires a certain initial 
velocity in order to become effective. The ion velocity has to exceed the 
Alfv\'en speed $V_\mathrm{A}=\mathrm{B_0}/\sqrt{4\pi\rho}$ while the
electrons must have  
velocities at least above $\sqrt{m_\mathrm{i}/m_\mathrm{e}}V_\mathrm{A}$. This has been called the
'injection problem'. 

2) {\sl Acceleration by parallel electric fields:}
Direct acceleration by electric fields depend on its strength compared to the 
Dreicer field $E_D=(2\pi e^3n_\mathrm{e}\ln{\Lambda})/(k_\mathrm{B}T)$. If 
$E>E_\mathrm{D}$ most electrons and ions gain energy. If
$E<E_\mathrm{D}$ only electrons 
in the high energy tail of the velocity distribution function will be
accelerated. The limitation in both cases is the maintenance of
overall neutrality of charge and pre-existing current in the
acceleration region.
 
3) {\sl MHD turbulence:} This acceleration mechanism occurs when particles 
interact many times with randomly moving MHD waves. Due to a slight overplus 
in head-on collisions the interaction results in an
energy gain for the particle. As in the shock acceleration model, the
acceleration by MHD turbulence suffers from an 'injection
problem': Thermal ions and electrons cannot resonate with MHD waves
for typical solar pre-flare conditions.

A solution for this problem is the assumption of MHD turbulent cascades
that channel the energy residing in the MHD turbulence to smaller scales and
into the region where interaction with thermal particles is
possible. A realization of this scenario is proposed
in Miller (\cite{miller1991}), Miller \& Roberts (\cite{millerroberts1995}) and
Miller (\cite{miller1997}). An MHD turbulent cascade transfers the energy
from large scale MHD waves to smaller scales where the energy may be
absorbed by the particles. The mechanism that  
dissipates the wave energy into the particles is transit-time 
damping (Fisk \cite{fisk1976}; Stix \cite{stix1992}). It is is basically a
resonant Fermi acceleration of second order. Only particles in
resonance with a low-amplitude MHD wave are affected. The resonance
condition is the usual $l=0$ (or Landau) resonance given by
$\omega-k_\parallel v_\parallel\approx 0$. As the particles can only  
gain energy in the direction parallel to the background magnetic field, the 
temperature in parallel direction increases.

A preference for acceleration along the background magnetic field is a
common feature of the acceleration models mentioned above. The 
velocity distribution thus becomes more and more anisotropic during
acceleration. If energization in parallel direction is from a thermal
level of some 0.1 keV to 20 keV or more but the perpendicular
temperature remains constant, the anisotropy is substantial. 
The free energy residing in parallel direction may give rise to growth
of plasma waves.
   
For $T_\parallel^\mathrm{e}>T_\perp^\mathrm{e}$ and high beta plasmas,
Hollweg \& V\"olk (\cite{hollwegvolk1970}) and Pilipp \& V\"olk
(\cite{pilippvolk1971}) have proposed the Electron Firehose
instability. This instability is an extension to higher frequencies
of the (MHD) Firehose instability, originally mentioned by Parker
(\cite{parker1958}). While the Firehose instability is of a completely
non-resonant nature, the Electron Firehose instability involves
non-resonant electrons but resonant protons.
 
For large anisotropy of the electron distribution, the electrons
become also resonant. This instability is described in
Pilipp \& Benz (\cite{pilippbenz1977}) and is called the Resonant
Electron Firehose instability.

Having been applied to a variety of problems, the Electron Firehose 
instability has not been considered to occur during electron acceleration in
solar flares.

Here we investigate the threshold for growth of plasma
modes resulting from acceleration and infer a
prediction for the evolution of the distribution function in velocity
space with respect to conditions expected to occur in solar flares.

We assume that no significant instability of Langmuir waves
occurs. This is suggested by the following argument: If a large
fraction of the available energy normally did go into Langmuir waves,
we would expect always a radio signature orders of magnitude higher
than ever observed during the impulsive phase (Benz \& Smith
\cite{benzsmith1987}). 

Section 2 describes the techniques used to solve the dispersion
equations. In the following section the results obtained
are shown and the thresholds of the instability are presented. In section 4 we 
discuss the effect on the acceleration of electrons and conclude this work. 
%------------------------------------------------------------------------------
%SECTION 2
%------------------------------------------------------------------------------
\section{Method}
Consider electromagnetic transverse waves of the form
$\exp{(ikx-i\omega t)}$ propagating in the direction of the background
magnetic field in an electron-proton plasma. The plasma dispersion
equation can then be written as 
\begin{eqnarray}\label{eqn:1}
\det\left( (\mbox{\bf 1}k^2-\vec{k}\vec{k})\frac{c^2}{\omega^2}-
\vec{\vec{\epsilon}}(\omega,\vec{k})\right)=0,
\end{eqnarray}
where, according to linearized kinetic theory, the dielectric tensor
$\vec{\vec{\epsilon}}(\omega,\vec{k})$ is given by 
\begin{eqnarray}\label{eqn:2}
&&\vec{\vec{\epsilon}}(\omega,\vec{k})=\nonumber\\
&&\mbox{\bf
1}-\frac{\omega_\mathrm{p}^2}{\omega^2}\Bigg\{\mbox{\bf
1}-\sum\limits_j\sum\limits_{n=-\infty}\limits^{\infty}
\int\vec{v}\vec{\vec{\prod}}\times\\ 
&&\frac{\frac{n\Omega_j}{v_\perp}\frac{\partial}{\partial
v_\perp}+k_\parallel\frac{\partial}{\partial v_\parallel}}{\omega
-k_\parallel v_\parallel -n\Omega_j}f^0_j\Bigg\}\nonumber.
\end{eqnarray}
The gyrofrequency of the $j$th species is given by $\Omega_j=q_jB/(cm_j)$
and $\omega_\mathrm{p}$ denotes the plasma frequency defined as
$\omega_\mathrm{p}=\sqrt{\sum_j\omega^2_{\mathrm{p}_j}}=
\sqrt{\sum_j 4\pi n_jq_j^2/m_j}$. The tensor $\vec{\vec{\prod}}$
is given by the matrix
\begin{eqnarray}\label{eqn:3}
&&\vec{\vec{\prod}}=\\
&&\left(
\begin{array}{ccc}
\left(\frac{n\Omega_j}{k_\perp}J_n\right)^2
&i\frac{n\Omega_j}{k_\perp}v_\perp J_nJ^\prime_n
&\frac{n\Omega_j}{k_\perp}v_\parallel J_n^2\\
-i\frac{n\Omega_j}{k_\perp}v_\perp J_nJ^\prime_n
&\left(v_\perp J^\prime_n\right)^2
&-iv_\perp v_\parallel J_n J^\prime_n\\
\frac{n\Omega_j}{k_\perp}v_\parallel J_n^2
&iv_\perp v_\parallel J_n J^\prime_n
&\left(v_\parallel J_n\right)^2
\end{array}
\right)\nonumber
\end{eqnarray}
where the argument of the Bessel function $J_n$ is $k_\perp
v_\perp/\Omega_j$. $f^0_j(v_\parallel,v_\perp)$ in equation
(\ref{eqn:2}) denotes the zero order distribution 
function in velocity space of the particle species $j$.
In order to obtain full solutions of this equation, the computer code
WHAMP (R\"onnmark \cite{roennmark1982}) has been applied to the
problem. The usage of this code has been facilitated by programming an
interface for the programming language IDL. It is called IDLWhamp
and provides the user with a comfortable tool to input parameters to
the code and for management and visualization of the results.

According to the capability of the WHAMP code, the most general form
of the particle distribution function 
$f^0_j(v_\parallel,v_\perp)$  for each species $j$ is given by
\begin{eqnarray}\label{eqn:4}
&&f^0_j(v_\parallel,v_\perp)=\nonumber\\
&&\frac{1}{\sqrt{2\pi}v_{j_\mathrm{th}}}
\exp{\left(-\left(\frac{v_\parallel}{2v_{j_\mathrm{th}}}-v_{dj}\right)^2\right)}
\times\\
&&\Bigg[\frac{\Delta_{j}}{\alpha_{1j}}\exp{\left(-\frac{v_\perp^2}
{2\alpha_{1j}v^2_{j_\mathrm{th}}}\right)}+\frac{1-\Delta_{j}}{\alpha_{1j}-
\alpha_{2j}}\times\nonumber\\ 
&&\Bigg\{\exp{\left(-
\frac{v_\perp^2}{2\alpha_{1j}v^2_{j_\mathrm{th}}}\right)}-\exp{\left(-\frac{v_\perp^2}
{2\alpha_{2j}v^2_{j_\mathrm{th}}}\right)}  \Bigg\}\Bigg]\nonumber.
\end{eqnarray}
This is the original notation used in R\"onnmark
(\cite{roennmark1982}) beside the choice 
of the thermal speed to be
$v_{j_\mathrm{th}}=\sqrt{(k_\mathrm{B}T_j/m_j)}$. The 
$\alpha_{1j}$ parameter is the  anisotropy
$\alpha_{1j}=T^j_\perp/T^j_\parallel$ of the $j$-th distribution
function. $\Delta_{j}$ and $\alpha_{2j}$ 
define the depth and size of a possible loss-cone.

We assume that the electron velocity distribution function  
can be described by a bi-maxwellian with different temperatures in
parallel and perpendicular direction with respect to the background
magnetic field. Hence for our problem the parameters $\Delta_{j}$ and
$\alpha_{2j}$ were set to unity in equation (\ref{eqn:4}).

Taking into account the uncertainties in the acceleration region
including a possible pre-heating mechanism altering
the pre-flare plasma conditions, we do not want to restrict  
our work to the parameters of a particular scenario. According
to Pallavicini et al. (\cite{pallaviciniseriovaiana1977}) reasonable
ranges in the acceleration region of an impulsive solar flare would be
$\approx 100-500\mathrm{G}$ for the background magnetic field, $\approx
10^9-10^{11}\mathrm{cm}^{-3}$ for the number density and $\approx
10^6-10^7\mathrm{K}$ for the temperature of electrons and protons.  
%------------------------------------------------------------------------------
%SECTION 3
%------------------------------------------------------------------------------
\section{Results}
%------------------------------------------------------------------------------
%SUBSECTION 3.1
%------------------------------------------------------------------------------
\subsection{Electron Firehose Instability}
\begin{figure}
 \resizebox{\hsize}{!}{\includegraphics{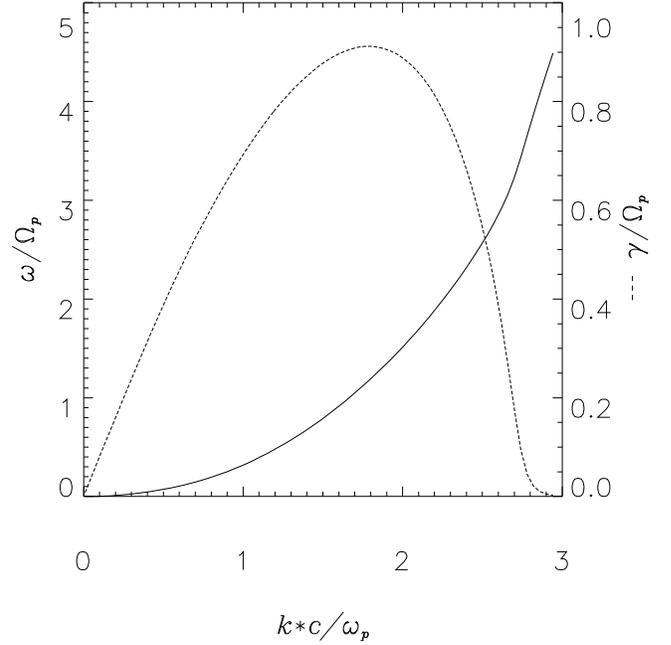}}
 \caption[]{A typical plot of the dispersion relation. The chosen
 parameters are $T^\mathrm{e}_\perp=T^\mathrm{p}_\perp=
 T^\mathrm{p}_\parallel=10^7\mathrm{K}$,
 $\frac{T_\parallel}{T_\perp}=20,n_\mathrm{e}=5\cdot 10^{10}
 \mathrm{cm}^{-3},\mathrm{B_0}=100\mathrm{G}$. The real part of the
 frequency $\omega_\mathrm{r}$ and the growth rate  $\gamma$ are
 normalized to the proton gyrofrequency $|\Omega_\mathrm{p}|$. The
 parallel wave vector is normalized to the proton inertial length. The
 whole branch is lefthand circularly polarized. }
 \label{fig:1}
\end{figure}

The only mode exhibiting significant growth rates in our
calculations is a lefthand circularly polarized wave which was
identified to be the Electron Firehose instability.
This mode evolves out of a stable righthand polarized whistler wave at
small anisotropy. With increasing anisotropy, the frequency 
$\omega_\mathrm{r}$ is shifted so that, with the convention $\omega_\mathrm{r}>0$, the
mode becomes lefthand circularly polarized at $k|| \mathrm{B_0}$ in the
unstable regime, cf. section 7 in Gary (\cite{gary1993}). A typical
dispersion relation is plotted in Fig.~\ref{fig:1}.  
%------------------------------------------------------------------------------
%FIGURE 1
%------------------------------------------------------------------------------
%------------------------------------------------------------------------------
%------------------------------------------------------------------------------
%FIGURE 2
%------------------------------------------------------------------------------
\begin{figure}
\resizebox{\hsize}{!}{\includegraphics{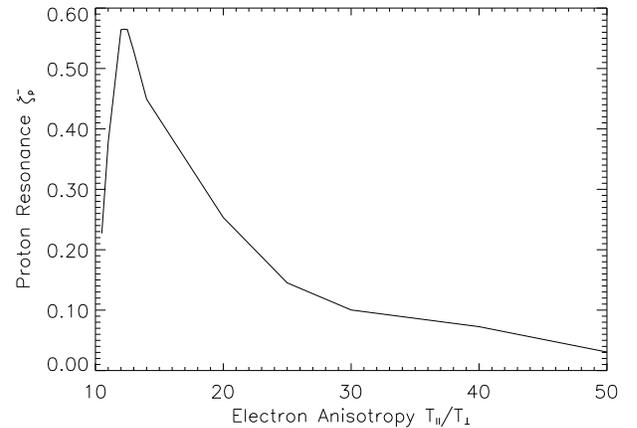}}
\caption[]{Resonance factor $|\zeta_\mathrm{p}^{-}|$ of the protons versus the
electron anisotropy $\frac{T_\parallel}{T_\perp}$ for the fastest
growing modes. The larger $|\zeta^-_\mathrm{p}|$, the smaller is the
fraction of protons in resonance.}
\label{fig:2} 
\end{figure}
%------------------------------------------------------------------------------
By introducing the resonance factor
\begin{eqnarray}\label{eqn:5}
\zeta^\pm_j\equiv\frac{\omega_\mathrm{r}\pm\Omega_j} 
{\sqrt{2}|k_\parallel|v_{j_\mathrm{th}}},
\end{eqnarray}
the values for the protons and electrons are found to be
$|\zeta_\mathrm{p}^{-}|\sim 1$ and $|\zeta_\mathrm{e}^{-}|\gg1$, 
demonstrating resonance for
the protons and non-resonance for the electrons.
 
According to Hollweg \& V\"olk (\cite{hollwegvolk1970}) there are also
right hand 
circularly polarized modes, which can become unstable for this extension
of the Firehose instability. These modes have been found, but the
growth rates are smaller than the growth rates of the lefthand
polarized modes described above.  

As the instability first appears, the phase velocities of the resonant
waves are near the peak of the proton distribution.
Fig.~\ref{fig:2} displays a plot of the proton resonance factor 
(\ref{eqn:5}) for the fastest growing modes versus electron
anisotropy. 
%------------------------------------------------------------------------------
%FIGURE 3
%------------------------------------------------------------------------------
\begin{figure}
\resizebox{\hsize}{!}{\includegraphics{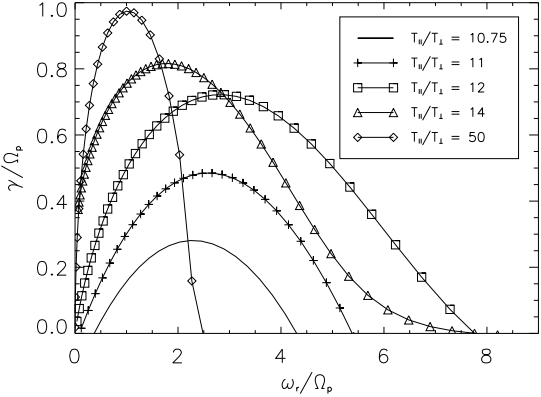}}
\caption[]{The growth rate $\gamma$ versus the real part of the
frequency $\omega_\mathrm{r}$, both normalized to the proton gyrofrequency, for
different anisotropies $\frac{T_\parallel}{T_\perp}$. The values of
the other plasma parameters are the same as in Fig.~\ref{fig:1}.}
\label{fig:3}
\end{figure}
%------------------------------------------------------------------------------
With increasing anisotropy less protons are resonant and the
resonance factor increases. The change in the fraction of resonant protons
is mirrored in the excited frequency range. As depicted in
Fig.~\ref{fig:3} the unstable frequency range grows to a maximum value
at an anisotropy of about $\frac{T_\parallel}{T_\perp}\sim 12$,
coinciding with the maximum value of the resonance factor at
$|\zeta_\mathrm{p}^{-}|\sim 0.57$ of the protons
(cf. Fig.~\ref{fig:2}). As the resonance factor decreases again, the
excited frequency range becomes narrow around
$\omega_r/|\Omega_\mathrm{p}|\sim1$. This narrow range is in itself
evidence for the resonant character of the instability.
%------------------------------------------------------------------------------
%SUBSECTION 3.2
%------------------------------------------------------------------------------
\subsection{Instability Threshold}
In this section we present the calculated threshold for linear growth
of L-mode waves excited by the Electron Firehose
instability.

The initial plasma is assumed to be maxwellian with temperatures
$T_{\perp 0}=T_{\parallel 0}=T_{0}$ perpendicular and
parallel to the background magnetic field for both plasma species, the
electrons and the protons. Taking into account an acceleration
mechanism for the electrons acting only in parallel direction, the
perpendicular temperature remains constant throughout the whole
acceleration process, i.e. $T^\mathrm{e}_\perp=T^\mathrm{e}_{\perp 0}$.  
In order to investigate the condition of the pre-flare plasma
for the Electron Firehose instability to occur during the acceleration
process, the initial plasma parameters have to be connected to the
actual plasma parameters during the acceleration. With the assumptions
above, this can be done by defining an initial parallel plasma beta,
$\beta^\mathrm{e}_{\parallel 0}$, via the perpendicular plasma beta
\begin{eqnarray}\label{eqn:6a}
\beta^\mathrm{e}_{\parallel0}\equiv\beta^\mathrm{e}_\perp=\frac{8\pi n_\mathrm{e} k_\mathrm{B} 
T^\mathrm{e}_\perp}{\mathrm{B_0}^2}, 
\end{eqnarray}
and the usual parallel plasma beta by
\begin{eqnarray}\label{eqn:6b}
\beta^\mathrm{e}_\parallel\equiv\frac{8\pi n_\mathrm{e} k_\mathrm{B} T^\mathrm{e}_\parallel}{\mathrm{B_0}^2}=
\beta^\mathrm{e}_{\parallel0}\cdot\frac{T^\mathrm{e}_\parallel}{T^\mathrm{e}_\perp}, 
\end{eqnarray}
where the connection between these two quantities is given by the
temperature anisotropy $T^\mathrm{e}_\parallel/T^\mathrm{e}_\perp$.

%------------------------------------------------------------------------------
%FIGURE 4
%------------------------------------------------------------------------------
\begin{figure}
\resizebox{\hsize}{!}{\includegraphics{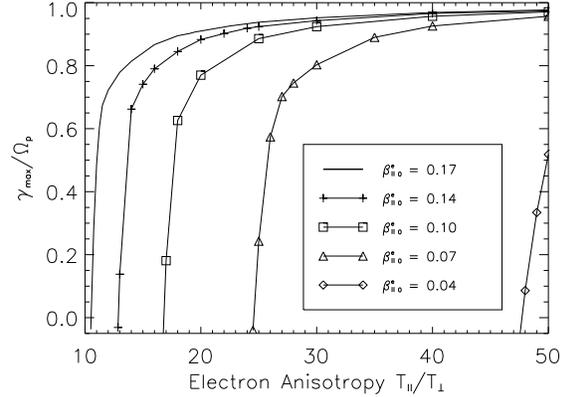}}
\caption[]{The maximum growth rate $\gamma_\mathrm{max}$ normalized to the
proton gyrofrequency $|\Omega_\mathrm{p}|$ versus the anisotropy
$T^\mathrm{e}_\parallel/T^\mathrm{e}_\perp$ of the electrons;
$T^\mathrm{p}_\perp=T^\mathrm{p}_\parallel$,
$T^\mathrm{e}_\perp=T^\mathrm{p}_\perp$. The appropriate frequency is
always of the order of $|\Omega_\mathrm{p}|$.} 
\label{fig:4} 
\end{figure}
%------------------------------------------------------------------------------
According to Hollweg \& V\"olk (\cite{hollwegvolk1970}) the instability
criterion for the Electron Firehose instability may be approximated by
\begin{eqnarray}\label{eqn:7}
1-\beta_\parallel^\mathrm{e}A_\mathrm{e}<0,
\end{eqnarray}
where the anisotropy factor is defined by
$A_\mathrm{e}=1-T^\mathrm{e}_\perp/T^\mathrm{e}_\parallel$. 

As one can see from inequality (\ref{eqn:7}), the instability
threshold does not depend directly on the parameters
$n_\mathrm{e},\;T^\mathrm{e}_\parallel,\;\mathrm{B_0}$,
but only on the resulting $\beta^\mathrm{e}_\parallel$. This
independence is also reproduced with the numerically obtained
data. For our purpose, the plasma is therefore fully described by the
plasma beta. 

In Fig.~\ref{fig:4} the function $\gamma_\mathrm{max}(T^\mathrm{e}_\parallel/ 
T^\mathrm{e}_\perp)$ is plotted for five different values of
$\beta^\mathrm{e}_{\parallel 0}$. The maximum growth rate of the instability
steeply raises at the threshold of the instability and flattens for
larger anisotropies, where $\gamma_\mathrm{max}/|\Omega_\mathrm{p}|$
approaches unity.    
 
\begin{figure}
\resizebox{\hsize}{!}{\includegraphics{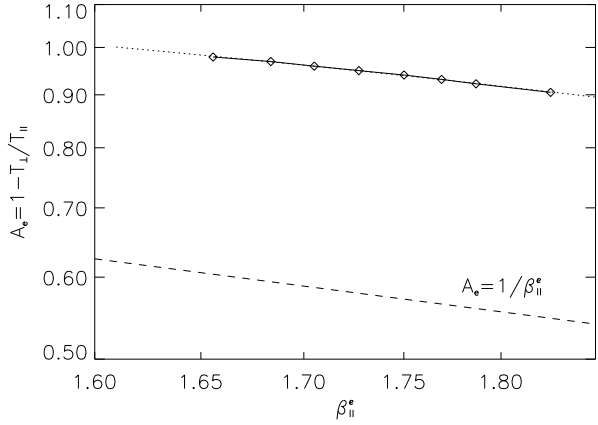}}
\caption[]{Threshold for the Electron Firehose instability in 
anisotropy factor vs. parallel electron beta. The scaling of
both axes is logarithmic. The dashed curve shows the
instability limit according to equation (\ref{eqn:7}). The dotted
line represents a fit to the numerically obtained values which are
depicted by diamonds. The areas above the respective lines are the
unstable regions.}
\label{fig:5} 
\end{figure}

From these results, the contour of zero growth rate
in the $A_\mathrm{e}-\beta^\mathrm{e}_{\parallel}$ plane has been derived
(cf. Fig.~\ref{fig:5}). The discrepancy between the analytically
derived relation (\ref{eqn:7}) and the numerically obtained values is 
due to the approximation used in Hollweg \& V\"olk (\cite{hollwegvolk1970}).
%------------------------------------------------------------------------------
%FIGURE 5
%------------------------------------------------------------------------------
%------------------------------------------------------------------------------
%------------------------------------------------------------------------------
%FIGURE 6
%------------------------------------------------------------------------------
\begin{figure}
\resizebox{\hsize}{!}{\includegraphics{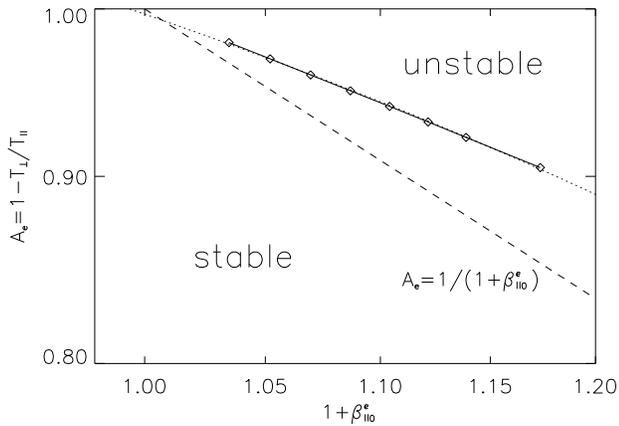}}
\caption[]{The same plot as Fig.~\ref{fig:5} but this time $A_\mathrm{e}$ has
been plotted vs. the initial electron beta. Again, as in 
Fig.~\ref{fig:5}, the dashed curve  
shows the instability limit according to equation (\ref{eqn:7}) and 
the dotted line represents a fit to the numerically obtained
values.}
\label{fig:6}   
\end{figure}
%------------------------------------------------------------------------------

According to inequality (\ref{eqn:7}), instability cannot occur for
a parallel beta smaller than unity. Due to the deviations
of the approximation mentioned above, this limit
is shifted to a value of $\sim 1.6$ (cf. Fig.~\ref{fig:5}). 

In order to investigate the necessary properties of the pre-flare 
plasma for the Electron Firehose instability to occur, it is
the initial plasma beta that is of interest. Fig.~\ref{fig:6}
depicts the same plot as Fig.~\ref{fig:5} but this time the anisotropy
factor $A_\mathrm{e}$ has been plotted versus the initial plasma beta
$\beta^\mathrm{e}_{\parallel 0}$. The dotted line in both figures
represents a fit to the numerically obtained values and is an
extrapolation to a broader range of beta values. The negative
$\beta^\mathrm{e}_{\parallel 0}$ at the $A_\mathrm{e}\rightarrow 1$
limit is an artifact of this extrapolation.   

The values of the initial plasma beta for the Electron Firehose
instability to occur at considerable values of
$T^\mathrm{e}_\parallel/T^\mathrm{e}_\perp$ are well within the range
of usually assumed pre-flare plasma parameters. For example, an
initial plasma beta of $\beta^\mathrm{e}_{\parallel 0}\approx0.05$ can be
realized by assuming pre-flare plasma parameters of 
$n_\mathrm{e}=5\cdot10^{10}\mathrm{cm}^{-3}$,
$T^\mathrm{e}_{0} = 3\cdot10^6K$ and $\mathrm{B_0} =
100\mathrm{G}$. This plasma becomes unstable at an anisotropy of
$T_\parallel/T_\perp\approx32$.    

According to the acceleration model via transit-time
damping, this is a reasonable value for the anisotropy to
occur during the acceleration process (Lenters \& Miller
\cite{lentersmiller1998}). 
%------------------------------------------------------------------------------
%SUBSECTION 3.3
%------------------------------------------------------------------------------
\subsection{Influence of Anisotropic Protons}
If we assume the protons to be heated by the same or a similar
mechanism, it is to be expected that they will grow anisotropic in the same
way the electrons do. Hence, we also have investigated the influence
of anisotropic protons and briefly discuss the effect of an additional
proton anisotropy on the instability.

Consider a plasma with anisotropic electrons and isotropic protons
that is already unstable to the Electron Firehose instability.
When the protons are anisotropized by increasing the temperature in
parallel direction, more and more 
become resonant to the L-waves, non-resonantly excited
by the electrons. As shown by Hollweg \& V\"olk
(\cite{hollwegvolk1970}), the protons are damping 
these waves. Hence, it is to be expected that the resulting   
growth rate of the L-waves decreases as the proton anisotropy is
increased. This expectation has been verified by numerical
calculation.

Moreover, the protons are heated by absorption of the excited waves at
the expense of the electrons (Pilipp \& V\"olk
\cite{pilippvolk1971}). According to Kennel \& Petschek
(\cite{kennelpetschek1966}) this scatters the protons to 
higher perpendicular velocities and hence, destroys or inverts the
parallel proton anisotropy.  It inhibits the growth of the electron
anisotropy and may complicate the acceleration to higher
energies, but increases the bulk energy of the protons. The energy
transfer from the electrons to the protons via the Electron Firehose
instability could be responsible for the proton energization, which is
a problem in the transit-time damping scenario (Miller
\cite{miller1998}).

If the protons are anisotropic, there is an additional righthand
polarized wave mode. This mode is the extension of the lefthand
Electron Firehose mode to negative frequencies. According to
Hollweg \& V\"olk (\cite{hollwegvolk1970}) 
this mode has real frequencies mainly below the proton
gyrofrequency and is in resonance with the protons. As the anisotropy
of the electrons increases, this righthand polarized mode becomes less and less
significant.  

The proton cooling through the instability of the righthand mode
competes with the heating by the lefthand mode and it is not yet clear
what the net energization of the protons will be.
%------------------------------------------------------------------------------
%SUBSECTION 3.4
%------------------------------------------------------------------------------
\subsection{Oblique Propagation: Preliminary Results}
%------------------------------------------------------------------------------
%FIGURE 7
%------------------------------------------------------------------------------
\begin{figure}[t]
\resizebox{\hsize}{!}{\includegraphics{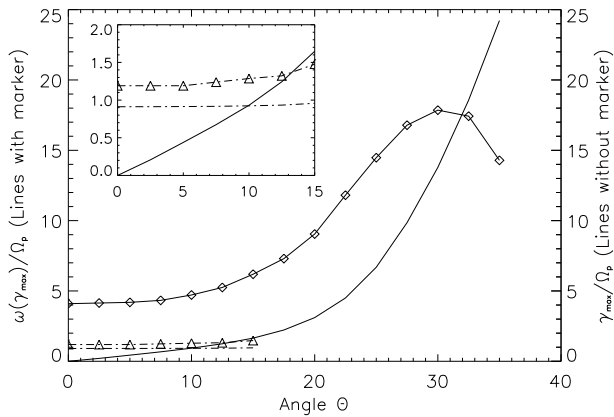}}
\caption[]{Frequency and the according growth rates vs. the
propagation angle $\Theta$ with respect to the background magnetic
field. The dashed lines indicate the branch of the parallel Electron
Firehose instability. The solid lines show the oblique mode that
exhibit much faster growth than the Electron Firehose instability. The
small plot is an enlargement of the region, where the crossing of the
growth rates occurs. The plasma parameters are the same as in
Fig.~\ref{fig:1}.} 
\label{fig:7} 
\end{figure}
%------------------------------------------------------------------------------
Sample calculations in oblique directions indicate an unstable branch
of solutions that grows faster than the parallel Electron Firehose
mode. This mode is stable at parallel propagation and is also
lefthand circularly polarized. Fig.~\ref{fig:7} depicts frequency and
maximum growth rate versus the angle $\Theta$ of both modes. The
dashed lines represent the branch, that is excited by the Electron
Firehose instability at parallel propagation. At a 
propagation angle of about $10^\circ$ with respect to 
the background magnetic field, the growth rate of the oblique mode
becomes larger than the growth rate of the parallel mode. Hence, the
determining mode for instability thresholds is the oblique mode
rather than the parallel Electron Firehose mode. 

As calculations have shown, not only the growth rate of the
oblique mode is larger than in the parallel case, but also the
instability threshold may be lower with respect to anisotropy. Plasmas
being stable with respect to the parallel mode exhibited instability
to the oblique mode. Therefore, the thresholds for instability derived
in the section above can be considered as upper boundaries.      

Lefthand circularly polarized oblique
modes seem to have never been considered in connection with the
Electron Firehose instability. The further investigation of these
modes is the subject of ongoing work.  
  
%------------------------------------------------------------------------------
%SUBSECTION 3.4
%------------------------------------------------------------------------------
\section{Discussion and Conclusion}
Numerical solutions of the dispersion equation for
lefthand circularly polarized electromagnetic waves, propagating
parallel to the background magnetic field, have shown that the Electron
Firehose instability, usually considered as a 'high-beta plasma' 
instability, must be expected in coronal plasmas which are processed by an
acceleration mechanism with a preference in parallel direction.
The distribution function of the electrons in velocity space
has been represented by a bi-maxwellian with temperatures $T^\mathrm{e}_\perp$
and $T^\mathrm{e}_\parallel$, perpendicular and parallel  with
respect to the background magnetic field. The protons have been
assumed to be isotropic and in thermal equilibrium. 
 
Considering the uncertainty in the pre-flare conditions, we have
investigated instabilities in a broad range of plasma temperature $T$, density
$n$ and background magnetic field $\mathrm{B_0}$. For these plasmas,
it was found that the Electron Firehose instability occurs at anisotropies
that must be expected for an acceleration mechanism acting
predominantly in parallel direction and being capable of producing the
observed electron bulk energization.
 
The unstable parallel modes that have been found are lefthand
circularly polarized and non-resonantly excited by the electrons, but
partially cyclotron resonant with the protons, which absorb the wave
energy. Hence, energy is transferred from the electrons to the
protons. Assuming the density and the magnetic field to be constant
during the acceleration process, there is a 
limiting electron temperature in parallel direction that cannot be exceeded
without loosing energy to the protons via the Electron Firehose
instability. The Electron Firehose instability may thus inhibit the
electron acceleration process and limit the reachable energies.

At angles $\Theta\not=0$ with respect to the background magnetic field,
an oblique mode has been found, that exhibits even larger growth rate than
the parallel Electron Firehose instability. This mode is also
lefthand circularly polarized. The properties of the  oblique mode
open new aspects on the Electron Firehose instability and its thresholds. 

Taking anisotropic protons into account, the lefthand mode may
extend to negative frequencies. They correspond to the righthand
circularly polarized mode resonantly excited by the
protons. Due to the cooling effect of this instability it is not yet
clear to what the net energization of the electrons and the protons will
amount. This topic will be the subject of future particle
simulations.

\begin{acknowledgements}
The authors thank S. Peter Gary and James A. Miller for their helpful
advice and Kjell R\"onnmark for providing them with a copy of the
KGI report describing the original WHAMP code. The authors also want
to acknowledge G\'erard Belmont and Laurence Rezeau who gave them free
access to their improved version of the WHAMP code, which has become
the mathematical core of IDLWhamp.

This work was financially supported by the Swiss National Science
Foundation (grant No. 20-53664.98).

\end{acknowledgements}

\end{document}